\documentclass[twocolumn,prb,aps,floatfix]{revtex4}
\usepackage{graphicx}
\usepackage{epsf}
\usepackage{bm}
\newcommand{\br}{{\bf r}}
\newcommand{\bS}{\vec{S}}
\newcommand{\bp}{\vec{p}}

\newcommand{\sR}{{\sf R}}
\newcommand{\sP}{{\sf P}}
\newcommand{\epij}{\epsilon_{ij}}
\newcommand{\half}{\frac{1}{2}}
\begin{document}

\title{Dirty quantum Hall ferromagnets and quantum Hall spin glasses}

\author{Derek K.K. Lee}
\affiliation{Blackett Laboratory, Imperial College, Prince Consort
Road, London, SW7 2BW, United Kingdom}

\author{Sebastian Rapsch} 
\author{J. T. Chalker} \affiliation{Theoretical Physics, University of Oxford,
1 Keble Rd, Oxford OX1 3NP, United Kingdom}

\date{\today}

\begin{abstract}
We study quantum Hall ferromagnets in the presence of a random
electrostatic impurity potential, within the framework of a classical
non-linear sigma model.  We discuss the behaviour of the system using
a heuristic picture for the competition between exchange and
screening, and test our conclusions with extensive numerical
simulations.  We obtain a phase diagram for the system as a function
of disorder strength, $\Delta$, and deviation, $\delta \nu$, of the
average Landau level filling factor from unity. Screening of an
impurity potential requires distortions of the spin configuration. In
the absence of Zeeman coupling there is a disorder-driven,
zero-temperature phase transition from a ferromagnet at small $\Delta$
and $|\delta \nu|$ to a spin glass at larger $\Delta$ or $|\delta
\nu|$. We characterise the spin glass phase in terms of its magnetic
and charge response.
\end{abstract}
\maketitle

\section{Introduction} 

Disorder and interactions have competing consequences in quantum Hall ferromagnets
(QHFM). In this paper, we study how a disordered impurity potential can give
rise to a spin-disordered ground state. We also discuss the
influence of disorder on the magnetic and charge response of such a
system. We use a classical spin model throughout
to describe the quantum Hall ferromagnet.

Coulomb interactions lead to spin correlations in a quantum
Hall system. For electrons fully occupying the lowest Landau level
(filling fraction $\nu=1$),
exchange is responsible for a spin-polarised ground state, even in
the absence of Zeeman energy. This is the consequence of Hund's rule
as applied to a macroscopically large number of degenerate
Landau-level orbitals. The resulting quantum Hall ferromagnet is especially
interesting as a system in which the spin configuration and the charge
density are closely linked \cite{leshouchesqhfm}. At $\nu=1$
and if Zeeman energy is large, charge
enters the spin-polarised system as minority-spin electrons.
However, if Zeeman energy is small or vanishing, the
charged excitation of lowest energy is not a bare spin-half electron,
but a bound state of an electron with many spin waves. In classical terms, this
occurs because the minority spin polarises its local ferromagnetic
background, and the composite object may be viewed as a topological excitation,
or texture in an ordered ferromagnet --- a
skyrmion\cite{sondhiskyrmion}. Similarly, an antiskyrmion, with
topological charge of the opposite sign, is produced when charge is
removed from a filled Landau level.
In this description, the deviation of
local charge density from that of a filled and ferromagnetically
polarised Landau level is proportional to the topological density
\cite{rajaraman} of the spin configuration.

In a clean system with sufficiently small Zeeman energy, skyrmions or
anti-skyrmions can be created at zero temperature on varying the
average filling factor from $\nu=1$ to larger or smaller values.
%They
%are also generated thermally in pairs at non-zero temperatures, along
%with spin waves.
For a disordered quantum Hall ferromagnet, the coupling of an
electrostatic impurity potential to the charge density offers an
additional mechanism by which spin textures are nucleated at zero
temperature. The physical consequences of this coupling are the subject of
this paper, which provides a detailed account of work
presented in outline in Ref.\,\onlinecite{rapsch02}.

The interplay between disorder and exchange in quantum Hall
ferromagnets has been examined previously from several different
viewpoints.  Fogler and Shklovskii \cite{foglershklovskii} have
developed a theory particularly applicable in higher Landau
levels. Building on earlier discussions \cite{ando73}, they set out
a mean-field treatment in the spirit of Stoner theory. For odd integer
filling in the absence of Zeeman coupling, they find a transition
between ferromagnetic and paramagnetic ground states with increasing
disorder strength. They suggest that this transition should be
apparent in transport measurements, in which the ferromagnetic phase
is characterised by spin-resolved Shubnikov-de Haas oscillations, and
the paramagnet by spin-unresolved oscillations. Experimentally, a
transition of this kind is observed with decreasing magnetic field
strength \cite{paalanen82,leadley98}, and its sharpness suggests that
its origin is indeed cooperative.

Within the Fogler-Shklovskii approach, local moments are all
collinear in the ferromagnet and vanish at the transition to the
paramagnet. An alternative scenario may arise in the lowest Landau
level near $\nu=1$, in which the QHFM responds to disorder mainly through
the direction rather than the magnitude of its local
magnetisation. Some indications that this can happen come from
calculations for the fully-polarised ferromagnet at weak
disorder. Here, a reduction in spin stiffness with increasing disorder
strength has been interpreted by Green \cite{green98} as a precursor
of a non-collinear phase. Moreover, even weak disorder may nucleate a
dilute glass of skyrmions and anti-skyrmions at the maxima and minima
of the disordered potential, as discussed by Nederveen and
Nazarov\cite{nazarov} and examined further in the present paper.
%(We will discuss further this scenario below.)
In addition, Sinova, MacDonald and Girvin \cite{sinova99} have shown
that, at intermediate disorder strength, both reduced and
non-collinear local moments emerge from a numerical solution of
Hartree-Fock theory for a model with Coulomb interactions and
spatially uncorrelated disorder, while transport properties within a
Hartree-Fock theory have been discussed by Murthy \cite{murthy01}.

In general, we believe that the relative importance of local moment
reduction versus the formation of spin textures for dirty QHFM's will
depend on the nature of the disorder. In this work, we
concentrate on textures, which are favoured by a smoothly varying
impurity potential at $\nu=1$. We find that the ferromagnet gives way
to a spin glass at strong disorder.

The plan of this paper is as follows. In the next section, we describe
a spin model with quenched disorder which is intended to capture the
physics of the disordered QHFM. This is followed in section
\ref{sec:phase} by a discussion of the phase diagram for the model as
a function of filling factor and disorder strength, using heuristic
arguments and scaling ideas. The conclusions are supported by the
results of a Monte Carlo simulation of a lattice version of the spin
model. The technique is outlined in section \ref{sec:mc} and the
results are presented in section \ref{sec:magnetic}. In sections
\ref{sec:dielectric} and \ref{sec:conduct}, we discuss the
compressibility and conductivity of the system in order to
characterise the charge response of spin-disordered ground states.

\section{\label{sec:model}Spin Model}

Consider a two-dimensional electron gas in a strong perpendicular
magnetic field $B$, with Landau-level filling $\nu$ close to
unity. The electrons are subject to an impurity
potential $V({\bf r})$ and an electron-electron interaction energy
U({\bf r}). 
%\begin{equation}
%E_{\rm I} = \frac{1}{2} \int \int 
%\rho({\bf r}) U({\bf r},{\bf r}') \rho({\bf r}')  d^2{\bf r} d^2{\bf r}'\,.
%\end{equation}
As a first step, let us omit the exchange interactions and the Zeeman
energy. Then the electron density $\rho({\bf r})$ is determined by
the balance between disorder and interactions, or in other words,
screening. We treat this using Thomas-Fermi theory. Such an
approximation has been applied by Efros\cite{efros88,efros89} to the
comparable problem in spin-polarised Landau levels when $\nu$ lies
near half-integer values. The ground-state charge density
$\rho({\bf  r})$ at weak disorder is determined by the condition that the
Hartree potential should match the chemical potential $\mu$
everywhere:
\begin{equation}
\mu=V({\bf r})+ 
\int  U({\bf r}-{\bf r}') \rho({\bf r}')   d^2{\bf r}'\,.
\label{screening}
\end{equation}
This approach is valid in the case where the resulting local filling
factor varies smoothly on the scale of the magnetic length,
$l_B=(\hbar/eB)^{1/2}$, and only has small fractional deviations from
$\nu=1$, so that
\begin{equation}
\delta \rho({\bf r}) \equiv \rho({\bf r})-(2\pi l_B^2)^{-1}\ll \rho({\bf r})\,.
\end{equation}

However, the Thomas-Fermi picture of good local screening does not take in
account exchange interactions. Provided electron density fluctuations are
small and smoothly varying,
ferromagnetic exchange leads locally to a maximal ferromagnetic
polarisation of the electron spins. This local magnetisation may vary in
space.  Denoting its direction by the three-component unit vector
$\vec{S}({\bf r})$, spatial fluctuations in spin orientation
are linked to electron
density by \cite{sondhiskyrmion,rajaraman}
\begin{equation}
\delta\rho({\bf r})= \frac{1}{8\pi}\; 
\epsilon_{ij}\, \vec{S}\cdot 
\left(\partial_i \vec{S}\times \partial_j \vec{S}\,\right)\,.
\label{rho}
\end{equation}
This direct connection is specific to the quantum Hall ferromagnet --- a
varying electron density implies a variation in the direction of the
local magnetisation and \emph{vice versa}. Such spin textures cost
exchange energy. So, a proper description of the system must include
exchange, impurity and Hartree contributions to the total energy of a
dirty quantum Hall ferromagnet. This brings us to study the
Hamiltonian \goodbreak
\begin{eqnarray}
 {\cal H} &=& \int
   \Big(\,\, \frac{J}{2} |\nabla\vec{S}({\bf r})|^2 + 
        [ V({\bf r}) - \mu ]\, \delta \rho({\bf r}) \nonumber\\
        &&\qquad\qquad +\, 
        \frac{U_0}{2}[\delta \rho({\bf r})]^2\,\, \Big)\,\, d^2{\bf r}\, ,
\label{model}
\end{eqnarray}  
where
\begin{equation}
J = \frac{1}{16 (2\pi)^{1/2}}\,\frac{e^2}{4\pi\epsilon_0\epsilon_r l_B}
\label{exchcoupling}
\end{equation}
is the exchange coupling\cite{moon95}. ($\epsilon_r$ is the relative
permittivity in the semiconductor.) At this point we have chosen for
simplicity a short-range Hartree interaction, $U({\bf r}-{\bf r}')=
U_0\delta({\bf r}-{\bf r}')$.  We have also absorbed the constant
$U_0/2\pi l_B^2$ into the chemical potential $\mu$.

As mentioned above, we will work with a disordered potential $V({\bf
r})$ that is smooth on the scale of the magnetic length $l_B$. For
simplicity, our discussion of this continuum model will assume a
Gaussian distribution with standard deviation $\Delta$ and correlation
length $\lambda$ much larger than the magnetic length $l_B$. (Our
numerical study will use a lattice model with a bounded distribution
with uncorrelated disorder.)  

In restricting our study to this model, we neglect quantum
fluctuations of $\vec{S}({\bf r})$. This semiclassical approximation
is justified for smooth variations, with $|\nabla \vec{S}({\bf
r})| \ll l_{\rm B}^{-1}$.
% so that
%$\langle V({\bf r}) \rangle = 0$ and $\langle V({\bf r})V({\bf r}')\rangle
%=\Delta^2 f(|{\bf r} - {\bf r}'|/\lambda)$, where $f(0)=1$ and 
%$f(x)=0$ for $x \gg 1$ (a more realistic treatment of the potential
%due to remote ionised donors in a modulation-doped heterostructure
%presents no additional problems of principle). 
Our aim in the following is to understand the zero-temperature phase
diagram of the model defined by Eq.\,(\ref{model}), as a function of
disorder strength $\Delta$ and average charge density $\langle \delta \rho
\rangle$, the spatial average of $\delta \rho({\bf r})$. We will
characterise its ground states via their response functions and
excitations.

We finish this section by comparing this model with some other
examples of disordered
systems. As an electron system, it is unusual
in that there is an exchange gap for single-particle excitations, even
if the ground-state spin configuration $\vec{S}({\bf r})$ does not
have long-range ferromagnetic order. This means that the only low-energy
excitations involve collective spin modes. As a
ferromagnet with quenched disorder, the system is also unusual in
several ways. First, the link between spin and charge means that the
spin system responds to applied electric fields. We calculate in the following the
wavevector-dependent dielectric constant, $\epsilon(q)$, and compare
it with behaviour found in more conventional disordered electron
systems. Second, due to the same coupling, spin waves generate an
electric dipole moment. This means that spin waves contribute to the
optical conductivity $\sigma(\omega)$.  More generally, the coupling
to disorder in this model leaves spin-rotational symmetry intact but
breaks time-reversal symmetry. This is in contrast to the effect of
random Zeeman fields, which break both symmetries, and to random
exchange interactions, which preserve both symmetries.

\section{\label{sec:phase}Phase Diagram}

We begin with a qualitative discussion of the phase diagram as a
function of disorder strength and average charge density.
We employ scaling arguments to obtain the phase boundary for
breakdown of long-range ferromagnetic order.

The Hamiltonian of Eq.\,(\ref{model}) for the continuum model is minimised by
the spin configuration which satisfies (appendix \ref{app:spinwave}):
\begin{eqnarray}
J \,\nabla^2\bS \times \bS &=&\epij \partial_i V_{\rm
H}(\br)\,\partial_j\bS\\
V_{\rm H}(\br) &=& V(\br)-\mu + U_0\delta\rho(\br)
\label{spingndstate}
\end{eqnarray}
where $V_{\rm H}$ the local Hartree potential and $\delta\rho$ is
defined in terms of the local spin configuration by Eq.\,(\ref{rho}).
This equation for the ground state configuration shows the interplay
between exchange and disorder in the model. Since it is difficult to tackle
the non-linear equation directly, we proceed using heuristic
arguments instead.

The model is characterised by two energy scales: the exchange energy
$J$ and the disorder strength $\Delta$. There are also two
length scales: the correlation length $\lambda$ of the disordered
potential and
\begin{equation}
L_{\rm H} \equiv (U_0/J)^{1/2}\,,
\label{hartree}
\end{equation}
which we call the Hartree length.
The significance of the Hartree length can be made clear by
considering a skyrmion of fixed shape and radius $R$ in a clean
system. The contribution to its total energy from exchange is
$\sim J$ and independent of size, while that from Hartree interactions is
size dependent, being $\sim U_0/R^2$. Comparing these contributions, one sees
that exchange dominates on length scales large compared with $L_{\rm H}$, while
Hartree interactions dominate at smaller distances.

Our central hypothesis is that the competition between interaction and
disorder in this system is characterised by the Hartree length only.
In the following, we also use the limit
%\begin{equation}
$L_{\rm H} \gg \lambda$
%\label{hartreelimit}
%\end{equation}
as a source of simplifications. 

\subsection{Filled Landau level}
Let us consider first the effect of disorder on a system at $\nu=1$,
imposed by setting $\langle \delta \rho \rangle = 0$.
Without impurities ($V({\bf r}) = 0$),
the system is a perfect ferromagnet.
Moreover, there is a threshold \cite{green98,nazarov,sinova99}
\begin{equation}
|V({\bf r})|= 4\pi J.
\label{threshold}
\end{equation}
below which an impurity potential is unscreened. It arises
because, for any $\vec{S}({\bf r})$, one has $|\nabla
\vec{S}({\bf r})|^2 \geq 8\pi |\delta \rho({\bf r})|$ and hence
\begin{equation}
{\cal H} \geq \int \left[4\pi J
|\delta \rho({\bf r})| + V({\bf r})\delta \rho({\bf r})\right] d^2{\bf r}
\end{equation}
Thus, if $|V({\bf r})|$ is below the threshold $4\pi J$
everywhere, the ground state is the perfectly aligned ferromagnet with
$\delta \rho({\bf r})=0$.

At weak disorder ($\Delta \lesssim J$), $|V({\bf r})|$ in most parts of the system lies
below the threshold. With an unbounded potential distribution, the
ground state spin configuration therefore consists of a
dilute glass of skyrmions and anti-skyrmions,
nucleated at rare positions where $|V({\bf r})|$ is large, as discussed by
Nederveen and Nazarov\cite{nazarov}. 
Away from these positions we expect that the ferromagnetic order is
essentially unaffected by the impurity potential. A careful treatment of
this regime is however quite subtle, since the spin deviation due to an isolated
skyrmion falls off with distance only as $r^{-1}$. We argue in 
appendix \ref{app:dilute}
that long range ferromagnetic order is indeed preserved, and that
the internal degrees of freedom
of dilute skyrmions and antiskyrmions develop the correlations necessary
to ensure this.

In contrast, at strong disorder ($\Delta \gg J$), the charge density
provides almost perfect local screening of the disordered potential,
so that
%, from Eq.\,(\ref{screening}),
\begin{equation}
\delta \rho({\bf r}) \simeq - V({\bf r})/U_0 \qquad (\Delta/J\gg 1)\,.
\end{equation}
Corrections to perfect Thomas-Fermi screening arise at length
scales larger than $L_{\rm H}$, where exchange becomes important.
The effect of exchange is to force screening charges to be
quantised, since an unquantised charge costs divergent exchange
energy in the thermodynamic limit. 
We can summarise the effect of exchange by dividing the system into
regions of area $L_{\rm H}^2$, finding for each such area the
integral
\begin{equation}
Q\equiv -\int_{L_H^2} \frac{V({\bf r})}{U_0}\, d^2{\bf r}
\label{screencharge}
\end{equation}
and adjusting the total screening charge within every region to the
integer value closest to $Q$. We argue that these integers are
predominantly zero in a ferromagnetic phase, and predominantly
non-zero in a phase without ferromagnetic order. To see this, consider
a well-ordered ferromagnetic phase, in which ${S}({\bf r})$ has
small spatial variations around a global direction of magnetisation.
In this case, the net topological charge in any region has magnitude
much less than one. Conversely, in a phase without such order, unit
topological charge will typically accumulate over a region of linear
size given by the ferromagnetic correlation length. 

This picture leads us to identify the phase boundary of the
ferromagnet as the point at which $\langle Q^2 \rangle^{1/2}\sim 1$.
To estimate $\langle Q^2 \rangle$, note from
Eq.\,(\ref{screencharge}) that each correlation area of size
$\lambda^2$ contributes to $Q$ a charge of magnitude $\lambda^2\Delta/U_0$ and random sign.
Over the Hartree area, there are $L_H^2/\lambda^2$ such
contributions, so that
\begin{equation}
\langle Q^2 \rangle^{1/2}\sim
(\lambda^2\Delta/U_0)(L_H/\lambda) = \lambda L_{\rm H}\Delta/U_0.
\end{equation}
The phase boundary is therefore given by
\begin{equation}
\Delta_c \sim U_0/(\lambda L_{\rm H}) = J(L_{\rm H}/\lambda).
\label{neutralscaling}
\end{equation}
For $\Delta > \Delta_c$, the ground state is strongly disordered and
has no ferromagnetic order. Within our classical description, the
spins are frozen at zero temperature. We can therefore identify this
phase as a quantum Hall spin glass.

\subsection{Away from integer filling}

Consider next the ground state away from $\nu=1$.  We examine
behaviour at fixed $\Delta < \Delta_c$, as a function of the average
charge density $\langle \delta \rho \rangle \propto \nu-1$. For
$\langle \delta \rho \rangle =0$ and $\Delta < \Delta_c$, the system
has a net magnetisation, but the introduction of charge, in the form
of skyrmions (or anti-skyrmions) disrupts ferromagnetic spin
alignment. The size $R$ of an isolated skyrmion in a clean system is
divergent if Hartree repulsion is not balanced by Zeeman
energy. However, the presence of an impurity potential establishes an
optimal size, because it is energetically favourable for a skyrmion to
locate its charge distribution in randomly occurring potential
wells. The typical depth of such a well of radius $R \gg \lambda$, 
averaged over its area, is 
$-\Delta(\lambda/R)$. (The case $\lambda > R$ is treated in
Ref.\,\onlinecite{nazarov}.)  The Hartree energy is $\sim
U_0/R^2$. The total energy is hence minimised if
\begin{equation}
R\sim U_0/(\Delta \lambda) = L_{\rm H}(\Delta_c/\Delta)\,.
\end{equation}
Note that this value of $R$ exceeds $L_{\rm H}$ for $\Delta <
\Delta_c$. Because of this, we expect that exchange energy dominates over the
potential energy and that the skyrmion will not be strongly distorted by the
random potential. In other words, screening as discussed in the
previous subsection does not alter the present argument.

We expect that ferromagnetic order will persist with
increasing charge density until such skyrmions overlap.
The phase boundary hence lies at
\begin{equation}
\langle \delta \rho \rangle_c= L_{\rm H}^{-2}(\Delta/\Delta_c)^2  \,.
\label{criticaldensity}
\end{equation}

\begin{figure}[htb]
\includegraphics[height=0.6\columnwidth,angle=270]{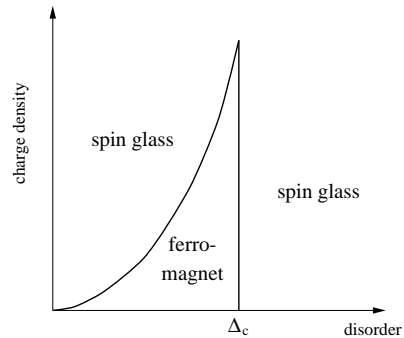}
\caption{Schematic diagram for $L_{\rm H}/\lambda\gg 1$.}
\label{fig:phasetheory}
\end{figure}

In summary, we have used simple arguments to obtain the phase
boundaries between the ferromagnetic and the spin glass as a function
of disorder and charge density. The results,
Eqs.\,(\ref{neutralscaling}) and (\ref{criticaldensity}), are summarised in
the schematic phase diagram shown in Fig.\,\ref{fig:phasetheory}. We
next present results from Monte Carlo simulations which support
these predictions.

The discussion we have presented is for a model with a short-range
Hartree interaction and for the limit $l_B \ll \lambda \ll L_{\rm H}$.
The central consequence of using a Coulomb form,
$U(r)=e^2/4\pi\epsilon_0 r$, in place of a short-range Hartree interaction
is to change the length scale derived by comparing Hartree with
exchange energies, from $L_{\rm H}\equiv(U_0/J)^{1/2}$ to 
$L_{\rm H}\equiv e^2/4\pi \epsilon_0 J = 16 (2\pi)^{1/2} l_B$. 
Since in this case $L_{\rm H}$ is not parametrically larger than $l_B$,
scaling arguments of the type presented in this section are
not justified. In this case, in place of three distinct regimes of 
disorder strength 
($\Delta \ll J$, $J\ll \Delta \ll \Delta_c$ and $\Delta_c \ll \Delta$)
we expect only two, with $\Delta_c \sim J$.

\section{\label{sec:mc}Monte Carlo simulations}

In order to test the theoretical results derived in the previous
section, we use Monte Carlo methods, in combination with simulated
annealing, to study a classical Heisenberg model with quenched
disorder. In this section, we outline our methodology. 
Similar techniques have been applied recently to systems without
quenched disorder in Ref.\onlinecite{hale00}.
We present our
results for the physical response functions of the model in sections
\ref{sec:magnetic} and \ref{sec:dielectric}.

We treat a lattice version of Eq.\,(\ref{model}) with $N
\times N$ classical spins $\vec{S}_i$ of unit length on a square
lattice, taking periodic boundary conditions.
Spins have nearest-neighbour ferromagnetic interactions of strength
$J$.
% The spins of our lattice model are taken to represent the local
%magnetic moments. In the limit of a smoothly varying disorder
%potential, the electron spins of the two-dimensional electron gas
%locally line up to form a large spin.  We replace this by a classical
%spin of fixed length.
In addition, the electrical charge $q_k$ on each plaquette $k$ is
calculated from the area on the spin sphere covered by the spins on
the corners of that plaquette. This charge density has a local
repulsive Hartree interaction of strength $U_0$, and is also subject
to a uniformly distributed random background potential
$\epsilon_k\in[-\Delta,\Delta]$. Since we choose this potential
independently for each plaquette, the correlation length $\lambda$ is
set by the lattice spacing.

The Hamiltonian of the lattice model is
\begin{eqnarray}
H&=& -J \sum_{\langle ij\rangle}
\vec{S}_{i}\cdot\vec{S}_{j}
+ \sum_{k}
%\left
(\epsilon_k q_k
+\frac{U_0}{2}q_k^2
%\right
)\nonumber\\
&&+\gamma
%\left
(Q_0-\sum_{k} q_k
%\right
)^2 \,,
\label{latticeham}
\end{eqnarray}
where we have introduced a
Lagrange multiplier $\gamma$ to bias the system towards a predefined
number $Q_0$ of charge quanta.

To obtain the ground state spin configuration for a given disorder
realisation, we start from a random initial state and anneal
using Monte Carlo dynamics and the Metropolis
algorithm. After some experimentation, we found the following three-stage
protocol to be effective. In the first stage, the temperature
is reduced linearly in time from a high temperature $T_0$ 
(several times $J$) to $T_0/10$,
using $3.5 \times 10^5$ Monte Carlo steps per spin (MCS).
In the second stage, the temperature is reduced from $T_0/10$ to $0$,
using $5 \times 10^5$ MCS. For both these stages, the attempted
spin update is an isotropically distributed reorientation.
In the third stage, the system is quenched for a further $5 \times 10^4$ MCS,
using as the attempted spin update only small angle reorientations
in order to improve
the acceptance rate.

We have checked that this algorithm finds the ground state reliably
for a weakly disordered system with overall charge neutrality, by
doing repeated runs for a given realisation of background potential
and using local charge and energy densities to identify states that
differ only by a global spin rotation.  In strongly disordered systems
and those with non-zero average charge density, repeated applications
of the algorithm do not reproduce the same state to high
precision. Instead, a number of low-lying states are obtained, having a
small spread (less than 10\%) in their energies and other observables
such as their magnetisation. 
We perform at least five independent simulations for every
disorder realisation and pick from the states obtained the one with
the lowest energy. For a chosen disorder strength $\Delta$, we generate
three or more realisations of the disorder potential and average our
results over these realisations.  We find that properties such as the
ground state energy, the magnetisation and various response functions
show only small fluctuations between different disorder realisations
for systems of size 32 $\times$ 32 spins or larger. Simulations of
very much bigger systems are ruled out by constraints on computing
time, and we choose $N=40$ as the system size for most of our
simulations.

The Lagrange multiplier $\gamma$ is chosen sufficiently large that
a majority of the simulation runs yield a ground state
with the desired total charge $Q_0$. We do not detect any dependence
of our results on the particular value of $\gamma$.

\section{\label{sec:magnetic}Magnetic Response}

Having outlined our simulation techniques, we next present our
results. We start with the magnetic properties of the system. We show
that they indicate a transition from the ferromagnet to the spin
glass with increasing disorder strength, as expected from our discussions in
section \ref{sec:phase}.

\subsection{Magnetisation}
\begin{figure}[hbt]
\includegraphics[width=\columnwidth]{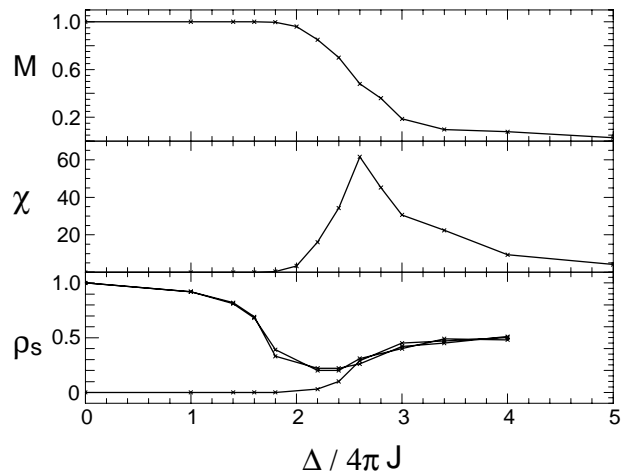}
\caption{Ground state properties for a system with $U_0=8\pi J$ as a
function of disorder strength: (a) magnetisation $M$ as a fraction of
the saturated moment, (b) susceptibility $\chi$, (c) spin stiffness
$\rho_s$ in units of $J$.}
\label{fig:mcdata}
\end{figure}

We calculate the site-averaged magnetisation
\begin{equation}
M=|\langle
\vec{S}({\bf r})\rangle|=N^{-2}|\sum_{i}\vec{S}_i|\,.
\end{equation}
Results are shown for a system of 40$\times$40
spins with $U_0=8\pi J$ in Fig.\,\ref{fig:mcdata}(a).
Spins are fully aligned for
$\Delta/4\pi J < 1$, because our bounded disorder distribution
then lies entirely below the threshold, 
Eq.\, (\ref{threshold}). Increasing $\Delta$
beyond $4\pi J$, we obtain a partially polarised ferromagnet. In fact,
the magnetisation remains close to its saturated value until
$\Delta/4\pi J \simeq 1.8$, when it starts to drop appreciably.
Increasing the disorder strength further, we reach a regime of
strong disorder at $\Delta/4\pi J \simeq 3$, in which there is only a small
magnetisation, of the magnitude expected for
a spin-glass state of a finite-size system.

These results are consistent with the existence of a phase transition from
the collinear, ordered phase at weak $\Delta$ to a spin-glass phase at
strong disorder. From the magnetisation curve we pick
$\Delta_c/4\pi J \simeq 2.5$ as our estimate of the critical disorder strength.

\subsection{Susceptibility}
We also calculate the uniform susceptibility $\chi$ from the
response of the ground state to a weak Zeeman field.
Our procedure is as follows. First we obtain a ground state without
a Zeeman field,
using the protocol described in Sec.\,\ref{sec:mc}.
Then we apply a weak Zeeman field $\vec{h}$ in the direction of the residual
magnetisation, by adding to the Hamiltonian the perturbation
$\delta H = - \vec{h}\cdot\vec{M}$.
We find the ground state in the presence of this perturbation
by repeating the third stage of the quenching protocol.
Taking care to check that our measurements remain within
the linear response regime, we extract
the susceptibility from the change in the magnetisation $\delta\vec{M}$, using
\begin{equation}
\chi=|\delta \vec{M}|/|\vec{h}|.
\end{equation}
Results for $U_0=8\pi J$ are shown in Fig.\,\ref{fig:mcdata}(b).
The susceptibility as a function of disorder strength has a large peak at
$\Delta/4\pi J=2.5$. We interpret this as a second indication of a phase
transition from the ferromagnetic phase to a disordered phase.

\subsection{Spin Stiffness}
We now turn to the spin stiffness $\rho_s$ which measures the rigidity
of the spin configuration. It is obtained by calculating the energy
cost of small amplitude, 
long-wavelength spin twists in the ground state. More specifically,
let the columns of sites in the $N\times N$ lattice be labelled by
integers $1, 2, \ldots N$. Then, starting with a ground state
obtained as in Sec.\,\ref{sec:mc}, we construct a twisted state
by rotating all spins on column $N/2$ of the system through a
small angle $\theta$ about an axis $\vec{e}$. Using the
third stage of the quenching protocol, we then relax all spins
in this twisted state except those on columns $1$
and $N/2$, which are held fixed.  From the difference
in energy $\Delta E$ between the initial and final states, we obtain the spin
stiffness for rotations about the axis $\vec{e}$, using
\begin{equation}
\rho_s=\Delta E/2\theta^2\,.
\end{equation}
Repeating this for different
axes of rotation, we calculate the full $3\times 3$ symmetric
tensor for the spin stiffness, $\rho_s^{\alpha\beta}$.
As expected, in the ferromagnetically ordered phase one of the
principal axes of this tensor lies to a good approximation
along the magnetisation direction, and it is convenient
in these calculations to choose rotation axes
$\vec{e}$ in directions parallel and perpendicular to  $\vec{M}$.

%We want to calculate the eigenvalues of the spin stiffness tensor.  We
%find that there is a natural choice of basis in spin space which
%approximately diagonalises this tensor. The first axis $\vec{e}_1$ is
%the direction of the net magnetisation. This should be a principal
%axis of $\rho_s^{\alpha\beta}$ due to rotation symmetry in spin space.
%This is also approximately the principal axis with the largest eigenvalue
%for the spin-spin correlation tensor:
%\begin{equation}
%A^{\alpha\beta} = \frac{1}{N^4} \sum_{ij} \left( S^\alpha_i S^\beta_j -
%\delta^{\alpha\beta}/3\right)
%\end{equation}
%To define as a second axis $\vec{e}_2$, we project the principal axis
%of $A$ with the second largest eigenvalue onto the plane perpendicular
%to the magnetisation. The third axis is chosen to be orthogonal to the
%first two.

The eigenvalues of the spin stiffness tensor are shown as a function
of disorder strength in Fig.\,\ref{fig:mcdata}(c).
For the fully-polarised ferromagnet, rotations
about the magnetisation direction do not alter the spin configuration,
and so one eigenvalue is zero while the other two are degenerate,
taking the value $\rho_s=J$.
% As expected, the fully-polarised ferromagnet at $\Delta=0$ has
%$\rho_s=J$.
For the partially-polarised ferromagnet, all three
eigenvalues are non-zero, with two remaining degenerate. The
spin-glass state, however, has no special spin
direction and this magnetic isotropy means that that all three
eigenvalues are approximately degenerate. The stiffness is reduced in
value from
$J$ in the fully polarised ferromagnet to approximately $J/2$ in the
spin glass. (A variational estimate for the stiffness
in the spin-glass phase is $2J/3$. See appendix \ref{app:spinwave}.)
The magnetically isotropic phase is observed for $\Delta/4\pi J > 2.5 $,
yielding the same estimate of $\Delta_c$ as our magnetisation and
susceptibility data.

\subsection{Spin Correlation Length}
The behaviour of spin correlations provides a further way of
characterising ground states. In particular, we consider
the correlation function
\begin{equation}\label{corrfn}
C(r)=\frac{1}{N_r}\sum_{(r)}\vec{S}_i\cdot\vec{S}_{j}\,,
\end{equation}
where the sum runs over all $N_r$ spin pairs of separation $r$. We
extract the spin correlation length $\xi$ by fitting to the form
\begin{equation}
C^\prime(r)=M^2+(1-M^2)\exp(-r/\xi)
\end{equation}

\begin{figure}[htb]
\includegraphics[width=0.7\columnwidth,angle=270]{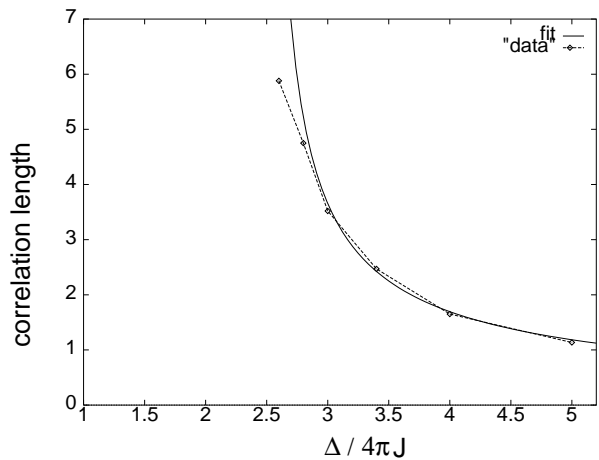}
\caption{Correlation length (in units of the lattice constant) as a
function of the disorder strength. Open symbols: our calculation,
solid line: fit to a power law.}
\label{fig5}
\end{figure}

The behaviour of the correlation length as a function of the disorder
strength in the spin-glass phase is shown in Fig.\,\ref{fig5}.  From
the Harris criterion\cite{harriscrit}, we expect it to diverge as $\xi
\sim (\Delta-\Delta_c)^{-\nu}$ with $\nu>2/d =1$ as the ferromagnetic
phase boundary is approached.  Our results are consistent with a
divergence at $\Delta/4\pi J = 2.5$, although it appears that our
they are affected by finite-size effects for $\xi$ for
$\Delta/4\pi J<2.7$. 
Perhaps because of these finite-size effects, this fit gives a low value
for the exponent: $\nu = 0.7$. Attempts at a similar fit in the ferromagnetic
phase are unsuccessful, and indeed the arguments of Appendix
\ref{app:dilute} suggest that correlations in this case may decay with
a power law.

%
%\vspace{2cm}

\subsection{Phase Diagram}

\begin{figure}[htb]
\includegraphics[width=0.9\columnwidth]{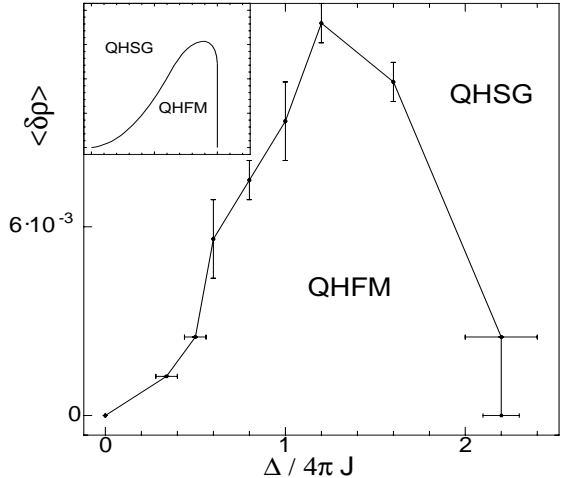}
\caption{The phase diagram for
  $U_0=4\pi J$ ($L_{\rm    H}/\lambda=\sqrt{4\pi}$),
  as a function of disorder strength and
  charge density. QHFM: quantum Hall ferromagnet, QHSG: quantum Hall
  spin glass. Inset: Results from scaling arguments for
  $L_{\rm H}/\lambda\gg 1$.}
\label{fig:phasemc}
\end{figure}

Our Monte Carlo results outlined above allow us to distinguish between
a ferromagnetic phase and a quantum Hall spin glass. For a system
with $U_0=8\pi J$, we conclude that a phase transition from a
collinear ferromagnet to a spin-disordered phase occurs
at $\Delta/4\pi J=2.5$.
Repeating these calculations for different disorder strength and charge
density, we can map out a phase diagram for the ground state
of the system. Results for $U_0=4\pi J$ are shown
in Fig.\,\ref{fig:phasemc}.

Comparing this with the phase diagram (Fig.~\ref{fig:phasetheory})
predicted from our heuristic arguments, we see that the two are
qualitatively very similar, even though the simulations are carried
out for $(L_H/\lambda)=\sqrt{4\pi}$ while our scaling picture applies
in the limit $(L_H/\lambda)\gg 1$. An idea of the dependence on
$L_H/\lambda$ is given by contrasting results at the two values of
$U_0$ studied.  The critical disorder $\Delta_c$ at which the neutral
system loses ferromagnetic order is $\Delta_c/4\pi J\simeq 2.5$ for
$U_0=8\pi J$, and reduces to $\Delta_c/4\pi J\simeq 2.2$ for $U_0=4\pi
J$, in qualitative agreement with the scaling behaviour we expect from
Eq.\,(\ref{neutralscaling}).  

It is interesting to note that there is a density range over which
disorder may \emph{stabilise} the ferromagnet, by limiting the size of
the nucleated spin textures. This range is however very narrow:
$|\langle \delta\rho \rangle| \lesssim 10^{-2}$ in units of charge per
plaquette.

%Again, we
%identify the transition in the neutral sector, in this case, at
%$\Delta/4\pi J=2.2$ from the fast decay of the magnetisation, a peak
%in the susceptibility and the fact that the three components of the
%spin stiffness tensor merge at that point indicating that a
%distinction between directions in spin space becomes
%obsolete. Evidence for the absence of long-range order in the spin
%glass phase comes from the correlation functions.
%

We mention in passing that we have not searched extensively for a
skyrmion crystal, expected at a finite charge density in the weak
disorder limit but presumably unstable for $\Delta \not=0$.

\section{\label{sec:dielectric}Dielectric response}

We have so far discussed the phase diagram of the system in terms of its
magnetic correlations and response. We next study its charge response.

We examine first the dielectric response of the partially polarised
ferromagnet and spin glass, characterised at zero frequency by the
wavevector-dependent dielectric constant $\epsilon(q)$ or by the
compressibility $\kappa(q)$, related via
\begin{equation}
\kappa(q)=q^2\epsilon(q)\epsilon_0/e^2\,,
\end{equation}
where $e$ is the electron charge and $\epsilon_0$ is the permittivity
of the medium. More precisely, we apply a periodic modulation, 
$V({\bf r}) \to V({\bf r}) + V_1 \cos({\bf q}\cdot {\bf r})$,
to the potential in
Eq.\,(\ref{model}). As a result, the ground-state electron density
changes according to $\delta \rho({\bf r}) \to \delta \rho({\bf r}) + \delta
\rho_1({\bf r})$. Since the system is disordered,
$\delta \rho_1({\bf r})$ contains many Fourier components, but after
averaging the linear response is
\begin{equation}
\langle
\delta \rho_1({\bf r})\rangle =
-V_1 \kappa(q)\cos({\bf q}\cdot {\bf r})\,,
\label{defcompress}
\end{equation}
which constitutes our definition of $\kappa(q)$.

\begin{figure}[hbt]
\includegraphics[width=0.8\columnwidth]{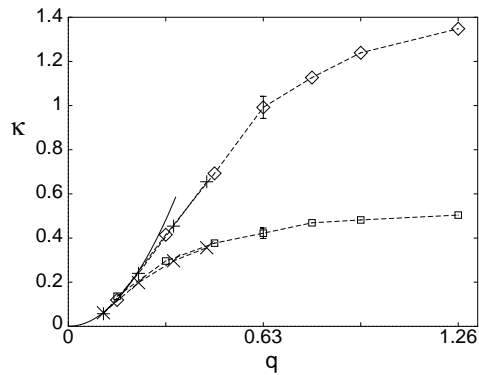}
\caption{Compressibility
  $\kappa(q)$ as a function of wavevector $q$ for systems with $J=1$
  and $U_0=4\pi$ and $8\pi$. ($\Delta/4\pi J=3$,
  $\langle\delta\rho\rangle=0$.)}
\label{fig:compress}
\end{figure}

Our numerical results for $\kappa(q)$ are displayed in
Fig.\,\ref{fig:compress}, where we study systems deep in the
spin-glass phase, setting  $\Delta/4\pi J=3$ and  
$\langle \delta \rho({\bf r})\rangle=0$.
We compare behaviours at $U_0/4\pi J=1$ and at $U_0/4\pi J=2$,
in each case combining data
from lattices of size $40^2$ and $56^2$ in order to
maximise wavevector resolution.

We find that, at small $q$, $\kappa(q)$ is independent of $U_0$ and
quadratic in $q$. At large $q$, it is independent of $q$, varying
roughly as $U_0^{-1}$. We can understand these results for $\kappa(q)$
using the approach we employed to discuss the phase diagram. The Hartree
length $L_{\rm H}$ again plays an important role.

For $q\gg L_{\rm H}^{-1}$, exchange may be
neglected and we see from Eq.\,(\ref{screencharge}) that
\begin{equation}
\kappa(q)\simeq U_0^{-1}\qquad (qL_H\gg 1)\,,
\end{equation}
in agreement with the $q$-independent value obtained for $\kappa(q)$
at large wavevectors from our simulations, and with
the $U_0$ dependence of these values.
\begin{figure}[hbt]
\includegraphics[height=0.8\columnwidth,angle=270]{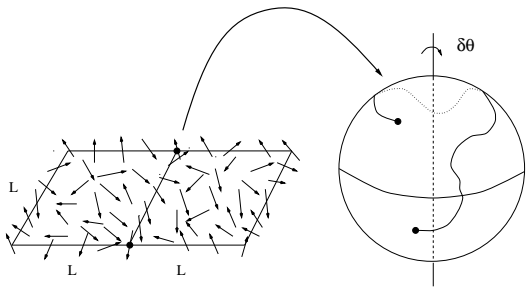}
\caption{Mapping a line of spins on the lattice to a trajectory in
spin space.}
\label{fig:map}
\end{figure}

Exchange becomes important at longer wavelengths $q\ll L_{\rm
H}^{-1}$.  To estimate $\langle \delta \rho_1({\bf r})\rangle$ 
for small $q$, we suppose that it arises primarily from spin
rotations which have amplitude $\delta\theta$
and wavevector $\sim {\bf q}$.
To be specific, consider for the periodic perturbation
$V_1\cos(q x)$
a region of size
$2L\times L$ with $L = \pi/q$. The perturbation causes a net movement of charge
$\delta Q_1$ in the $x$ direction, from one side of this region to the other.
We choose to focus on those spins lying on the line of length $L$
which is parallel to the $y$ axis and which divides
the region into two equal halves of size $L\times L$. (See
Fig.~\ref{fig:map}.) We suppose, in a variational spirit, that
the charge movement induced by the perturbation involves simply a
rigid rotation of these spins, through an angle $\delta\theta$.
With this assumption, let us estimate the charge transfer
generated by such a rotation, and the exchange energy that it costs.
To do so, it is useful to regard
the spin configuration as a map of the line we
have defined onto a trajectory on the
spin sphere. For a disordered spin configuration, this trajectory resembles a
random walk. The end points of the walk are randomly
placed on the spin sphere if $L$ is large compared to
the spin correlation length $\xi$, and the distance $D$ between them
on the surface of the spin sphere is therefore typically ${\cal O}(1)$.
Under a rigid rotation of spins on the line,
the trajectory is displaced rigidly around the spin sphere.
The charge transferred across the real-space
line is proportional to the area swept out on the spin sphere
by the trajectory during
this displacement. We therefore arrive at the estimate
\begin{equation}
\delta Q_1 \sim D\, \delta\theta \sim \delta\theta\,.
\end{equation}
Choosing the phase and axis of rotation appropriately in each such region,
the change in potential energy density arising from the rotation is
$ -V_1 \delta Q_1/L^2 = -V_1q^2\delta\theta$ while the associated
change in exchange energy density is $J q^2\delta\theta^2$. Choosing
$\delta\theta$ to minimise the total energy, we find $|\delta
\rho_1({\bf r})|\sim V_1q^2/J$ and hence
\begin{equation}
\kappa(q)\sim q^2/J = U_0^{-1} (qL_{\rm H})^2 \qquad(qL_H\ll 1)
\label{kappasmallq}
\end{equation}
This conclusion is again consistent with our numerical results.

Alternatively, we can arrive at this form for the compressibility
from scaling considerations. 
In general, we may expect that the compressibility is described 
by the scaling form
\begin{equation}
\kappa(q) = U_0^{-1} f(qL_H)\,,
\end{equation}
where $f(x)$ approaches a constant at large $x$. Our central
hypothesis is that exchange dominates at small wavevectors, and so
$\kappa(q)$ should be independent of $U_0$ as $q\to 0$. This implies
that $f(x)\sim x^2$ for $x\ll 1$, leading to the form given in
Eq.\,(\ref{kappasmallq}) for the compressibility at small $q$.

To summarise, the system has a metallic response to a perturbing
potential at large wavevectors, with $\kappa(q)$ independent of
$q$. However, at small wavevectors, it behaves like an insulator, with
$\epsilon(q)$ independent of q.

We note that, for an infinite system at exactly $q=0$ and zero
frequency, $\kappa(0)$ should be proportional to the thermodynamic
density of states.  So, except in the fully-polarised ferromagnet, we
expect that $\kappa(0)$ remains finite as we take the temperature to
zero. To reconcile this general expectation with our results, one must
remember that for finite-size systems considered here at zero
temperature, the discreteness of charge implies that almost all
disorder realisations have no response to an infinitesimal change in
the chemical potential, while there is a divergent response from those
realisations for which ground states from two different charge sectors
are degenerate.

\section{\label{sec:conduct}Optical Conductivity}

Finally, we consider the optical conductivity $\sigma(\omega)$ at
frequency $\omega$. Within our treatment, spin waves are the only
excitations that contribute. Since they are not topological
excitations and so do not carry net electrical charge, the dissipative
conductivity vanishes in the low-frequency limit.  Spin waves do,
however, give rise to local charge fluctuations and a fluctuating
electric dipole moment which couples to an oscillating external
electric field, generating dissipation at finite frequency.

For a fully-polarised quantum Hall ferromagnet, Green\cite{green99}
has shown that the spin-wave contribution to optical conductivity is
very small. In the non-collinear quantum Hall spin glass, the
contribution may be larger due to the presence of a finite charge
density in the disordered ground state. Furthermore, the low-energy
dynamics of the collinear quantum Hall ferromagnet is qualitatively
different from that of a quantum Hall spin glass, since while spin
waves in a collinear background have a quadratic energy dispersion,
those in a non-collinear background have a linear
dispersion\cite{halperinsaslow}, $\omega=cq$ at small $q$, with
velocity
\begin{equation}
c=(\rho_s/\chi)^{1/2}.
\label{spinwavevelocity}
\end{equation}
Of the three polarisation modes, one is
expected to remain gapless even in the presence of a Zeeman coupling.

We now calculate the spin-wave contribution to the optical
conductivity for a ground state with non-collinear spins, and then
estimate its magnitude in realistic systems.

The rotation of spins from their ground-state orientation
in the presence of a spin wave may be parameterised by
a vector ${\vec p}({\bf r})$, with at first order
% with wavevector ${\bf q}$ rotates the spins from their
%ground state by a small amount:
\begin{equation}
\vec{S} \rightarrow \vec{S} - \vec{p}({\bf r})\times\vec{S}  \,.
%\, ;
%\qquad
%\vec{p}({\bf r}) = \alpha \vec{p}_0 \cos({\bf q}\cdot{\bf r})
\label{smallrotation}
\end{equation}
%
%where $\vec{p}_0$ is a unit vector defining the axis of rotation.
Using Eq.\,(\ref{rho}), we show in appendix \ref{app:spinwave} that
this induces a change in the electron density
\begin{equation}
\delta \rho({\bf  r}) \to \delta\rho({\bf  r})
-\frac{\epsilon_{ij}}{4\pi}\,\partial_i\vec{p}\cdot\partial_j\vec{S}\,.
\label{spinwavecharge}
\end{equation}
%to leading order in $|\vec{p}|$. (See appendix
%\ref{app:spinwave}.)
This linear dependence on $\nabla \vec{p}$ should
be contrasted with spin waves in a collinear ferromagnet, for which the
density change is $O(|\nabla \vec{p}|^2)$.
%If the influence of ground state disorder on spin wave modes is neglected,
%one has $\vec{p}({\bf r}) = \alpha \vec{p}_0 \cos({\bf q}\cdot{\bf r})$,
%where $\vec{p}_0$ is a unit vector defining the axis of
%spin rotations.

To estimate the contribution to the conductivity, we note that a spin
wave will couple to an external electric field through the net dipole
moment ${\bf P}$ that it induces. The coupling appears in the
Hamiltonian in the form
\begin{equation}
\delta H = - P_x {\cal E}_x
\end{equation}
for an electric field ${\cal E}_x$ in the $x$-direction.
From Fermi's golden rule, the power $\Pi(\omega)$
absorbed from an oscillating electric field of frequency $\omega$ is
\begin{eqnarray}
\Pi(\omega)
%&=&
%\frac{2\pi}{\hbar^2}
%|\langle f | \delta H |0 \rangle|^2\, \hbar\omega\,g(\omega)
%\nonumber\\
&=& \frac{2\pi}{\hbar^2} {\cal E}_x^2
|\langle f | P_x |0 \rangle|^2\, \hbar\omega\,g(\omega)
\label{fermirule}
\end{eqnarray}
where $|0\rangle$ is the ground state and $|f\rangle$ is a state with
a single spin wave excited, of excitation energy $\hbar\omega$, and
$g(\omega)$ is the spin wave density of states in frequency. From
Eq.\,(\ref{spinwavevelocity}), we have
\begin{equation}
g(\omega)=\frac{3\omega}{2\pi c^2} L_xL_y \,,
\end{equation}
including three polarisations for a system with linear dimensions
$L_x$ and $L_y$.

We estimate for the matrix element appearing in Eq.\,(\ref{fermirule})
by considering the dipole moment induced by a spin wave.  We start our
discussion using the form taken by a spin wave mode in the absence of
ground-state disorder, $\vec{p}({\bf r}) = \alpha \vec{p}_0 \cos({\bf
q}\cdot{\bf r})$, where $\vec{p}_0$ is a unit vector defining the axis
of spin rotations and $\alpha$ specifies the amplitude.  The charge
density induced by an excitation of this type in a disordered ground
state with spin correlation length $\xi$ has, from
Eq.\,(\ref{spinwavecharge}), magnitude $\alpha q/\xi$.  We expect this
to fluctuate with a random sign over the length scale $\xi$. The
spin wave therefore induces electric dipoles of magnitude $e \xi
(q\xi\alpha)$ in each correlation area $\xi^2$.  Averaging over a
$L_x\times L_y$ system, we find
\begin{equation}
\langle P^2_x \rangle \sim
\left( e\xi^2 q \alpha \right)^2 \frac{L_xL_y}{\xi^2}
\sim e^2\alpha^2  \left(\frac{\omega\xi}{c}\right)^2 L_xL_y.
\end{equation}
where we have substituted $q=\omega/c$.

%We can therefore obtain a classical expression for the power:
%\begin{equation}
%\Pi(\omega)
%\sim \alpha^2 (\Delta V)^2 \frac{2\pi e^2}{\hbar^2}
%\left(\frac{\omega\xi}{c}\right)^2
%\hbar\omega\,\frac{3\omega L^2_y}{2\pi c^2}
%\end{equation}
%
%\begin{equation}
%\Delta E = - {\cal E}_x \langle P_x^2\rangle^{1/2} = \sim
%-|\vec{p}_0|\Lambda_q\, \Delta V
%\end{equation}
%where $\Lambda_q=e\xi q (L_y/L_x)^{1/2}$ is the coupling consta
%

It remains to determine the amplitude $\alpha$ for a single quantum
excitation. We show in appendix \ref{app:hydro} that
\begin{equation}
\langle \alpha^2\rangle = \hbar/\chi\omega L_xL_y  \,.
\label{zeropoint}
\end{equation}
Combining factors and dropping numerical coefficients, the absorbed power is
\begin{equation}
\Pi(\omega) \sim
{\cal E}_x^2 \frac{e^2\xi^2\chi}{\rho_s^2}\omega^3 L_x L_y \,.
\end{equation}
This is an ohmic response,
$\Pi/L_x L_y = \sigma {\cal E}^2_x/2$, with
conductivity
\begin{equation}
\sigma(\omega) \sim \frac{e^2}{h}\left(\frac{\omega \xi}{c}\right)^2
\frac{\hbar \omega}{\rho_s}
\end{equation}

Let us estimate the magnitude of this spin wave conductivity. 
%We combine the results from our numerical calculations with assumptions
%for typical values of the system parameters, taken for
%$GaAs$ at 10 tesla.
%\mbox{\AA}$. 
%To justify the assumption that the disorder varies slowly
%on scales of the typical electron separation, we further assume that
%$\xi=100 l_B$. 
In the spin glass phase, we use our numerical results to estimate
$\rho_s \simeq J/2$.  Also, the numerical results for the lattice
shows that $\chi_{\rm lattice}$ of the order of unity in units of
$l_B^2/J$. The continuum magnetisation is related to the lattice spins
by: $\vec{m} \leftrightarrow \vec{S}\hbar/l_B^2$. This means that the
continuum susceptibility is $\chi\simeq \hbar^2/Jl_B^2$.  Combining
these factors, we have
\begin{equation}
\sigma(\omega) \sim \frac{e^2}{h}\left(\frac{\xi}{l_B}\right)^2
\left(\frac{\hbar \omega}{\rho_s}\right)^3\,.
\end{equation}
Taking $\xi = 10 l_B$ and $J=$ 4-8 kelvin
(correcting our earlier value\cite{rapsch02}), 
we find for a frequency of 1 GHz,
$\sigma(\omega) \approx$ ($10^{-3}$-$10^{-4}$) $e^2/h$.
Unfortunately, variable-range hopping\cite{polyshklprb93} seems likely
to mask this contribution to $\sigma(\omega)$.

\section{Summary and discussion}

We have investigated the competition between exchange interactions and
disorder in quantum Hall ferromagnets at or near integer filling and
at zero temperature. Our approach is tailored to the limit of a
smoothly-varying impurity potential.

We find that the ferromagnetic state is destroyed by strong disorder
through the creation of skyrmion/anti-skyrmion pairs, or by a finite
density of either skyrmions or anti-skyrmions at filling factors
sufficiently far from $\nu=1$. This behaviour, anticipated from simple
scaling arguments, is confirmed in numerical studies of ground state
spin configurations, obtained from slow Monte Carlo cooling of an
initially random high temperature phase. The disordered phase is
identified as a quantum Hall spin glass by the absence of long-range
order, the presence of non-vanishing local magnetic moments and a
finite spin stiffness.

The quantum Hall spin glass has a zero-frequency dielectric response
which interpolates between that of an insulator at small wavevectors,
and that of a metal at large wavevectors. It supports gapless spin wave
modes that couple to electric fields through a finite dipole moment
and contribute to the optical conductivity of the system.

Possible experimental signatures of the phenomena we have discussed
follow both from the behaviour of the magnetisation
and from the nature of excitations. Measurements of the
Knight shift in nuclear magnetic resonance 
(NMR)\cite{barrett95,barrett01} provide
information on the distribution, sampled in space, 
of the spin polarisation component 
parallel to the applied magnetic field, 
while polarisation-resolved
absorption spectroscopy \cite{aifer96,nicholas02}
can be used to determine the average spin polarisation.
Impurity effects are most characteristic at $\nu=1$, where
they result in a reduced spin polarisation, reaching zero in the
spin glass at zero Zeeman coupling, and a finite width to the
polarisation distribution. We note that spin polarisation
which remains unsaturated even at $\nu=1$ is found in
absorption spectroscopy,\cite{aifer96,nicholas02} 
and that a broad Knight shift distribution
is measured in low-temperature NMR.)\cite{barrett01}
In addition, the existence
of excitations at energies lower than the Zeeman gap is
characteristic of the partially polarised ferromagnet,
and has been advanced as an explanation for a
coupling observed in NMR experiments
between radio-frequency magnetic fields
and the electron system.\cite{barrett98}
On the other hand,
at filling factors away from $\nu=1$, neither local
probes nor the nature of excitations distinguish
sharply between a skyrme crystal in a clean system
and the partially polarised ferromagnet or spin glass induced by disorder.

We are grateful for discussions with N. R. Cooper and S. L. Sondhi.
The work was supported in part by the EPSRC under
Grant GR/J78327 (JTC), and by the Royal Society (DKKL).

\appendix

\section{\label{app:spinwave}Spin waves and charge fluctuations}

Consider a small rotation $\sR(\bp)$ of the spins around $\bp$, so
that spin directions transform according to $\bS \to \sR\bS$, with
\begin{equation}
 \sR = e^{\sP} \qquad{\rm and} \qquad [\sP]^{ab} = \epsilon^{abc}p^c   \,.
\end{equation}
To second order, $\delta\bS \simeq -\bp\times\bS +\bp\times(\bp\times\bS)/2$.  The
charge density deviation from $\nu=1$ is given by $\delta\rho_0 = \epij
\bS\cdot(\partial_i\bS\times\partial_j\bS)/8\pi$. In the presence of an additional spin rotation,
it becomes
\begin{equation}
\delta\rho_0 + \delta\rho_1 = \frac{\epij}{8\pi} \sR\bS\cdot
\left[\partial_i(\sR\bS)\times\partial_j(\sR\bS)\right]
\end{equation}
so that
\begin{equation}
\delta\rho_1
%&=& \half \epij \bS\cdot
%        \left[2(\sR^{-1}\partial_i\sR) \bS \times \partial_j\bS +
%         (\sR^{-1}\partial_i\sR)\bS \times (\sR^{-1}\partial_j\sR)\bS\right]
%        \nonumber\\
= -\frac{\epij}{4\pi} \partial_i\bp\cdot\partial_j\bS +
        \frac{\epij}{8\pi} \partial_i\bp\cdot\partial_j(\bp\times\bS) + O(p^3)\,,
\end{equation}
obtained using $(\sR^{-1}\partial_i\sR)\bS \simeq (\partial_i\sP -
[\sP,\partial_i\sP]/2)\bS$ $=$ $\bS \times \partial_i\bp\, +$ $
(\partial_i\bp\times\bp)\times\bS/2$. We can drop higher-order terms
%for $\delta\rho_1$
if $|\bp|\ll 1$ and $|\nabla\bp| \ll |\nabla\bS|$, so
that $\delta\rho_1 / \delta\rho_0 \ll 1$.

To first order in $p$, the component of $\bp$ parallel to $\bS$ does
not affect the charge density. At this order, the charge fluctuation
$\delta\rho_1$ can be written as
\begin{equation}
  \delta\rho_1 = -\epij\partial_i\bp\cdot\partial_j\bS.
\end{equation}
From continuity, $\delta\dot\rho_1 +
\partial_i J_i = 0$, the current density is
\begin{equation}
J_i = \epij\dot{\bp}\cdot\partial_j\bS \quad\mbox{+ divergence-free part}
\label{current}
\end{equation}
Note that we have only identified the transport current, and
not any circulating current in the bulk.
%Therefore, this would not give the
%correct $\sigma_{xy}$.

Let us now consider the energy cost of spin rotations for the
Hamiltonian in Eq.\,(\ref{model}).  The change in the exchange energy
density $H_J$ is
\begin{eqnarray}
\delta H_J &=& J\,\partial_i\bp\cdot (\partial_i\bS \times \bS) +
        \delta H_J^{(2)}\,,\nonumber\\
\delta H_J^{(2)} &=& \frac{J}{2}\Big[\,|\nabla\bp|^2
- 2(\nabla\bp\cdot\bS)^2 +
(\partial_i\bp\cdot\bS)\partial_i(\bp\cdot\bS) \nonumber\\
&&\quad -  (\bp\cdot\bS)(\partial_i\bp\cdot\partial_i\bS)\,\Big]\,.
\end{eqnarray}
The change in potential energy density is
\begin{equation}
  \delta H_\rho = V_{\rm H}(\br) \delta\rho_1(\br)
  +\half\int\!\! \delta\rho_1(\br) U(\br-\br')\delta\rho_1(\br')d^2\br'
\end{equation}
where
\begin{equation}
V_{\rm H}(\br) = V(\br) -\mu + \int\! U(\br-\br')\delta\rho_0(\br')d^2\br'
\end{equation}
is the local Hartree potential and $\mu$ is the chemical potential.

An equation satisfied by the spin configuration in the ground state is obtained from
requiring that the
total energy is unaffected by the rotation $\vec{p}$ to first
order. This means that we have to balance the first-order terms in the
expressions for $\delta H_J$ and $\delta H_\rho$, giving
\begin{equation}
\partial_i\left[ J \,\partial_i\bS \times \bS - \epij V_{\rm
H}(\br)\partial_j\bS \right] = 0
\end{equation}
for the spins in the ground state. This non-linear equation
demonstrates the competition between ferromagnetic exchange and
Thomas-Fermi screening in a disordered QHFM.

From the second-order contributions $\delta H_J^{(2)}$ in the exchange energy
$\delta H_J$, we can make a variational estimate of the spin stiffness in the
spin-disordered phase. Suppose that $\vec{p}$ and $\vec{S}$ are uncorrelated.
Then the second-order terms average to
\begin{eqnarray}
 \langle(\nabla\bp\cdot\bS)^2\rangle
%&=& \frac{1}{3} |\nabla\vec{p}|^2 \nonumber\\
= \langle(\partial_i\bp\cdot\bS)\partial_i(\bp\cdot\bS)\rangle &=&
\frac{1}{3} |\nabla\vec{p}|^2 \nonumber\\
\langle(\bp\cdot\bS)(\partial_i\bp\cdot\partial_i\bS)\rangle &=& 0\,.
\end{eqnarray}
We can define a disorder-averaged spin stiffness from
$\langle\delta H_J^{(2)}\rangle = \tilde\rho_s |\nabla\vec{p}|^2/2$ so that
\begin{equation}
        \tilde\rho_s = 2J/3 \,.
\end{equation}

\section{\label{app:hydro}Long wavelength spin waves}

In this appendix, we review the results of Halperin and
Saslow\cite{halperinsaslow} for hydrodynamic
excitations,  and of Ginzburg\cite{ginzburg78} for
elementary excitations, and adapt these for our purposes.

Both theories deal with long-wavelength, low-energy
excitations of a disordered spin system. The microscopic magnetisation
density and energy density can
be coarse-grained over areas $A$ to mean values;
in the case of the magnetisation one has
\begin{equation}
 \vec{m}({\bf r}) = \frac{1}{A} \sum_{i\in A} \vec{S}_i  \,.
\end{equation}
Within a hydrodynamic theory, the low-energy dynamics of the system is
determined entirely these coarse-grained quantities, which obey
conservation laws and are assumed to fluctuate slowly on the timescale
set by local relaxation rates.

Hydrodynamic spin fluctuations have equations of motion which may be
written in terms of the rotation angle $\vec{p}$, as used in in
Eq.\,(\ref{smallrotation}), and the local magnetisation, with the form
\begin{eqnarray}
 \frac{\partial p_\alpha}{\partial t} = \chi^{-1} m_\alpha\\
 \frac{\partial m_\alpha}{\partial t} = \rho_s \nabla^2 p_\alpha
\end{eqnarray}
where $\rho_s$ is the spin stiffness and $\chi$ is the uniform
magnetic susceptibility.
Using the same variables, the free energy of the system is
\begin{eqnarray}
\Delta F(\vec{m},\vec{p})&=& {1\over 2} \sum_{\alpha=1}^{3} \int d^2 r
\left( \frac{1}{\chi} m_\alpha^2 + \rho_s |\nabla p_\alpha|^2\right)\nonumber\\
&=& {1\over 2} \sum_{\alpha=1}^{3} \int d^2 r
\left( \chi \dot{p}_\alpha^2 + \rho_s |\nabla p_\alpha|^2\right)   \,.
\end{eqnarray}
A long-wavelength treatment of elementary excitations leads
to the same equations of motion and an equivalent expression
for the energy of the system, in which
$\rho_s$ and $\chi$ are ground-state quantities.
We have argued in section \ref{sec:magnetic}
that both are finite at zero temperature in the spin-glass phase. These
equations are reminiscent of the dynamics of the spin waves in a
Heisenberg antiferromagnet. They lead to spin waves with the dispersion
relation
\begin{equation}
\omega=cq \, ;\qquad c = (\rho_s/\chi)^{1/2} \,.
\end{equation}

Considering a single spin-wave mode, with
\begin{equation}
\vec{p} = \alpha \vec{p}_0\cos({\bf q}\cdot{\bf r}-\omega t)
\qquad (|\vec{p}_0|=1) \,,
\end{equation}
we obtain the amplitude $\alpha$ for one quantum by setting the energy
to $\hbar \omega$, arriving at Eq.\,(\ref{zeropoint}).

%is a simple harmonic oscillator with ``mass'' $M=\chi L_x L_y/2$ and
%``spring constant'' $K=\rho_s q^2 L_x L_y/2$. We quantise this mode so
%that its contribution to the Hamiltonian becomes:
%\begin{equation}
%        H_{\rm sw} = \hbar\omega ( a^\dag a + {1 \over 2})
%\end{equation}
%where $a^\dag$ and $a$ are the creation and annihilation operators for
%this spin wave. Its amplitude $\alpha$ is given by:
%\begin{equation}
%        \alpha \leftrightarrow
%        \sqrt{\frac{\hbar}{2M\omega}} ( a^\dag + a) =
%        \sqrt{\frac{\hbar}{\chi\omega L_xL_y}} ( a^\dag + a)
%\end{equation}

\section{\label{app:dilute}Dilute Skyrmions}

In this appendix, we study in more detail the interactions between a dilute
set of skyrmions at weak disorder. We show how the relative
orientations of internal degrees of freedom
of the skyrmions are determined by the energetics of the system.

For a clean system without Hartree interactions, the Hamiltonian
reduces to the O(3) non-linear sigma model. The skyrmions are of the
Belavin-Polykov type. They do not interact. Hartree interactions alone
will lead to a divergent skyrmion size in the absence of a Zeeman
field. This is prevented in the presence of a disordered potential,
as pointed out in section \ref{sec:phase}. However, the presence of an
inhomogeneous potential also means that the skyrmions interact.  We
discuss here whether a physical picture of isolated skyrmions
at weak disorder is justified. In particular, since the spin deviation due
to a skyrmion falls off as $1/r$ with distance $r$, one should ask
whether the ferromagnetic polarisation is strongly affected by a
collection of dilute skyrmions.

For the purposes of this appendix, we choose a simple disorder
distribution in which pinned skyrmions and anti-skyrmions are
nucleated by isolated wells and barriers of circular
shape,
%with radius $\lambda$
positioned randomly on the
plane. The depth (or height) of these potentials exceeds the threshold,
Eq.\,(\ref{threshold}), so that they nucleate skyrmions (or
anti-skyrmions). This model disorder distribution has the advantage that we can use the
Belavin-Polyakov solutions of the non-linear sigma model as a starting
point for our analysis.

Consider first the case in which the chemical potential is sufficiently
high that there are no anti-skyrmions. A set of Belavin-Polyakov
skyrmions is described by an analytic function $w(z)$ of the position
$z=x+iy$ in the plane
\begin{equation}
        w(z) = \prod_{i=1}^n \frac{z-a_i}{z-b_i} \,,
\end{equation}
where $n$ is the number of skyrmions.
The function $w(z)$ parametrises the spin configuration via
\begin{eqnarray}
        S_x + iS_y &=& 2w/(1+|w|^2)\nonumber\\
        S_z &=& (1-|w|^2)/(1+|w|^2)
\end{eqnarray}
with the boundary condition $S_x=1$ at infinity.
% and $S_x=-1$ at the origin.
For a single skyrmion ($n=1$), the density profile is
\begin{equation}
\delta\rho_{n=1}(z) \equiv \frac{|\partial_z w|^2}{\pi(1+|w|^2)^2} =
\frac{d_i^2/4\pi}{(|z-z_i|^2 + d_i^2/4)^2}
\end{equation}
where $z_i=(a_i+b_i)/2$ can be interpreted as the position of the
skyrmion and $d_i = |a_i-b_i|$ as its diameter. There
remains an internal phase degree of freedom $\phi_i$, defined from $(a_i-b_i) =
d_i\exp(i\phi_i)$.
In
multiple skyrmions configurations the
relative phases are important.

In a dilute glass, skyrmions are typically separated by distances
large compared to their size, and the position and diameter of each skyrmion
is determined separately by the balance of Hartree and potential energy
in the vicinity of the potential well on which it is centered.
In the region far from any well, the Belavin-Polyakov form is a
good description for the spin configuration, because it minimises exchange energy.
The parameters $z_i$ and $d_i$ are fixed by the spin configuration
within the $i$th well, while the values of $\phi_i$ remain to be determined.
The density profile in this region is
\begin{equation}
\delta\rho'(z) = \frac{|w|^2}{\pi(1+|w|^2)^2}
        \left\vert \sum_i\frac{a_i-b_i}{(z-a_i)(z-b_i)}
        \right\vert^2
\label{multidensity}
\end{equation}
where the prime indicates the restricted
area in which this form applies.
The interference terms in the squared sum make
a contribution to the total charge in the region far from
the wells which amounts to a pairwise coupling between the internal
degrees of freedom of different skyrmions. The integrated density is
\begin{eqnarray}
I &=& \int'\frac{d^2z}{\pi}\,\, \frac{|w|^2}{(1+|w|^2)^2}\times\nonumber\\
        &&\mbox{Re}\sum_{i\neq j}
        \frac{(a_i-b_i)^*(a_j-b_j)}{(z^*-a^*_i)(z^*-b^*_i)(z-a_j)(z-b_j)}\quad
\end{eqnarray}
In the dilute limit, we can take $a_i\simeq b_i = z_i$. We can
also approximate $|w|^2\simeq 1$ away from the skyrmions. Then
\begin{eqnarray}
I &\simeq& \mbox{Re} \int'\frac{d^2z}{4\pi}\, \sum_{i\neq j}
        \frac{d_i d_j e^{i(\phi_i-\phi_j)}}{(z^*-z^*_i)^2(z-z_j)^2}
        \nonumber\\
&=& \frac{1}{2} \sum_{i>j} \frac{d_i d_j \cos(\phi_i-\phi_j)}{|z_i-z_j|^2}
\end{eqnarray}
This contribution to the charge in the region between potential wells
adds to the energy of the system, since the electrostatic potential
here does not exceed threshold [Eq.\,(\ref{threshold})].
It represents charge density that is removed from the cores of skyrmions,
as a result of overlap with the tails of other distant skyrmions. 
%The exchange energy cost is approximately $4\pi J |I|$.  This also means that the
%charge in the potential wells has changed by $-I$. This also raises
%the energy of the system by $\sum_i [ |V \Delta q_i| + U' (\Delta
%q_i)^2/2]$, where $\Delta q_i$ is the change in the charge in each
%well ($I=-\sum_i\Delta q_i$). $V$ is the potential in the well and
%$U'\propto U_0$ is an effective inverse capacitance.
We see that in order to minimise the energy  
we should choose the skyrmion phases
$\phi_i$ so as to minimise $I$. 
%, at least for small $U'$,
%we should minimise the interference contribution $I$ to the charge.
This is done by arranging $\phi_i$ so that the variable $d_i
e^{i\phi_i}$ sums locally to small values.   With such correlations,
we expect that long-range ferromagnetic order survives in the presence of
dilute skyrmions. 

%$\spadesuit$ It seems to me now
%that we {\it can} assume that the screened potential in the wells makes it
%favourable to transfer charge there, as is done when we minimise $I$. 
%If it were not so, then it would be better not to have the skyrmion in 
%the system at all. So I have left out part of your discussion. $\spadesuit$

A similar treatment of a mixed system containing both skyrmions and
antiskyrmions leads to interactions of the same kind amongst the
skyrmions, and separately amongst the antiskyrmions, but without
coupling between the two species at leading order in inverse density.

\bibliographystyle{apsrev} \bibliography{skyrme}

\end{document}